\documentclass[reprint,superscriptaddress,amsmath,amssymb,prl]{revtex4-1}
\usepackage[pdftex]{graphicx}
\usepackage{comment}
\usepackage{url}
\usepackage{bm}
\usepackage{comment}
\usepackage{setspace}
\usepackage{color}

\begin{document}

\title{Downward Terrestrial Gamma-ray Flash observed in Winter Thunderstorm}

\author{Y.~Wada}
\affiliation{Department of Physics, Graduate School of Science, 
    The University of Tokyo, 7-3-1 Hongo, Bunkyo-ku, Tokyo 113-0033, Japan}
\affiliation{High Energy Astrophysics Laboratory, Nishina Center for Accelerator-Based Science, 
    RIKEN, 2-1 Hirosawa, Wako, Saitama 351-0198, Japan}
\author{T.~Enoto}
\affiliation{The Hakubi Center for Advanced Research and Department of Astronomy,
    Kyoto University, Kitashirakawa Oiwake-cho, Sakyo-ku, Kyoto, Kyoto 606-8502, Japan}
\author{K.~Nakazawa}
\affiliation{Kobayashi-Maskawa Institute for the Origin of Particles and the Universe, 
    Nagoya University, Furo-cho, Chikusa-ku, Nagoya, Aichi 464-8601, Japan}
\author{Y.~Furuta}
\affiliation{Collaborative Laboratories for Advanced Decommissioning Science, Japan Atomic Energy Agency, 
    2-4 Shirakata, Tokai-mura, Naka-gun, Ibaraki 319-1195, Japan}
\author{T.~Yuasa}
\affiliation{Block 4B, Boon Tiong Road, Singapore 165004, Singapore}
\author{Y.~Nakamura}
\affiliation{Kobe City College of Technology, 8-3, Gakuen-Higashimachi, Nishi-ku, Kobe, Hyogo 651-2194, Japan}
\author{T.~Morimoto}
\affiliation{Faculty of Science and Engineering, Kindai University, 3-4-1 Kowakae, Higashiosaka, Osaka 577-8502, Japan}
\author{T.~Matsumoto}
\affiliation{Department of Physics, Graduate School of Science, 
    The University of Tokyo, 7-3-1 Hongo, Bunkyo-ku, Tokyo 113-0033, Japan}
\author{K.~Makishima}
\affiliation{Department of Physics, Graduate School of Science, 
    The University of Tokyo, 7-3-1 Hongo, Bunkyo-ku, Tokyo 113-0033, Japan}
\affiliation{High Energy Astrophysics Laboratory, Nishina Center for Accelerator-Based Science, 
    RIKEN, 2-1 Hirosawa, Wako, Saitama 351-0198, Japan}
\affiliation{Kavli Institute for the Physics and Mathematics of the Universe, 
    The University of Tokyo, 5-1-5 Kashiwa-no-ha, Kashiwa, Chiba 277-8683, Japan}
\author{H.~Tsuchiya}
\affiliation{Nuclear Science and Engineering Center, Japan Atomic Energy Agency, 
    2-4 Shirakata, Tokai-mura, Naka-gun, Ibaraki 319-1195, Japan}

\date{\today}

\begin{abstract}

During a winter thunderstorm on 2017 November 24, a strong burst of gamma rays with energies up to $\sim$10~MeV
    	was detected coincident with a lightning discharge, by scintillation detectors installed
    	at Kashiwazaki-Kariwa Nuclear Power Station at sea level in Japan.
    	The burst had a sub-second duration, which is suggestive of photoneutron productions.
    	The leading part of the burst was resolved into four intense gamma-ray bunches, 
    	each coincident with a low-frequency radio pulse.
    	These bunches were separated by 0.7--1.5~ms, with a duration of $\ll$1~ms each.
    	Thus, the present burst may be considered as a ``downward" terrestrial gamma-ray flash (TGF),
    	which is analogous to up-going TGFs observed from space.
    	Although the scintillation detectors were heavily saturated by these bunches, 
    	the total dose associated with them was successfully measured by ionization chambers,
    	employed by nine monitoring posts surrounding the power plant.
    	From this information and Monte Carlo simulations, the present downward TGF is suggested to have
    	taken place at an altitude of 2500 $\pm$ 500~m, involving $8^{+8}_{-4} \times 10^{18}$ 
    	avalanche electrons with energies above 1~MeV. This number is comparable to those in up-going TGFs.

\end{abstract}
\pacs{Valid PACS appear here}

\maketitle


\section{Introduction}
Terrestrial gamma-ray flashes (TGFs) are sub-millisecond gamma-ray emissions coincident with lightning discharges.
	Since their discovery in 1991\cite{Fishman_1994},
	TGFs have been observed by gamma-ray astronomy satellites\cite{Smith_2005,Marisaldi_2010,Tavani_2011,Briggs_2010,Briggs_2011,Mailyan_2016},
	and by aircrafts\cite{Smith_2011b,Bowers_2018}.
	The TGF photons, reaching 20~MeV or higher\cite{Smith_2005,Tavani_2011}, 
	originate from bremsstrahlung by energetic electrons.
	These electrons are thought to be accelerated by the relativistic runaway electron avalanche 
	(RREA) mechanism\cite{Gurevich_1992,Dwyer_2004b} in strong electric fields of lightning,
	and multiplied via such processes as the relativistic feedback processes\cite{Dwyer_2012a} 
	or the leader-seeded processes\cite{Celestin_2011,Babich_2015}.

Ground-level observations of natural\cite{Tran_2015,Colalillo_2017,Abbasi_2017,Abbasi_2018}
    	and rocket-triggered\cite{Dwyer_2004,Hare_2016} lightning have discovered gamma-ray flashes, similar to TGFs but beamed downward.
    	These bursts, called downward TGFs, indeed have characteristics common with ordinary TGFs,
    	such as coincidence with lightning, sub-millisecond duration, and photon energies of $>10~{\rm MeV}$.
	Recently, several downward TGFs in winter thunderstorms have been found to be so intense
	as to cause atmospheric photonuclear reactions, 
	such as $^{14}{\rm N}~+~\gamma~\to~^{13}{\rm N}~+~{\rm n}$\cite{Bowers_2017,Smith_2018,Enoto_2017}.
	
Since downward TGFs take place close to the ground, 
	brighter ones with high gamma-ray fluxes can heavily saturate detectors\cite{Hare_2016}.
	In particular, those bright enough to produce observable numbers of photoneutrons 
	completely exceeded counting capabilities of typical scintillation detectors\cite{Bowers_2017,Enoto_2017,Smith_2018}.
	As a result, we have not been able to estimate the gamma-ray fluence of downward TGFs that involve photonuclear reactions.
	Here we report on the detection, with scintillation detectors, of a downward TGF which triggered photonuclear reactions, 
	and estimate the gamma-ray fluence using ionization chambers tolerant to high-flux radiation.
    
\section{Instruments}
We have been performing the Gamma-Ray Observation of Winter Thunderclouds (GROWTH) experiment 
	in coastal areas of Japan Sea\cite{Tsuchiya_2007,Tsuchiya_2011,Tsuchiya_2013,Umemoto_2016,Enoto_2017,Wada_2018}.
	The present observation was performed in Kashiwazaki-Kariwa Nuclear Power Station, one of the GROWTH's sites. 
	This site, successfully operated since December 2006, was enhanced in 2016, with three gamma-ray detectors (Detectors A--C),
	of which the locations are shown in Fig.1.

\begin{figure}[t]
	\begin{center}
	\includegraphics[width=\hsize]{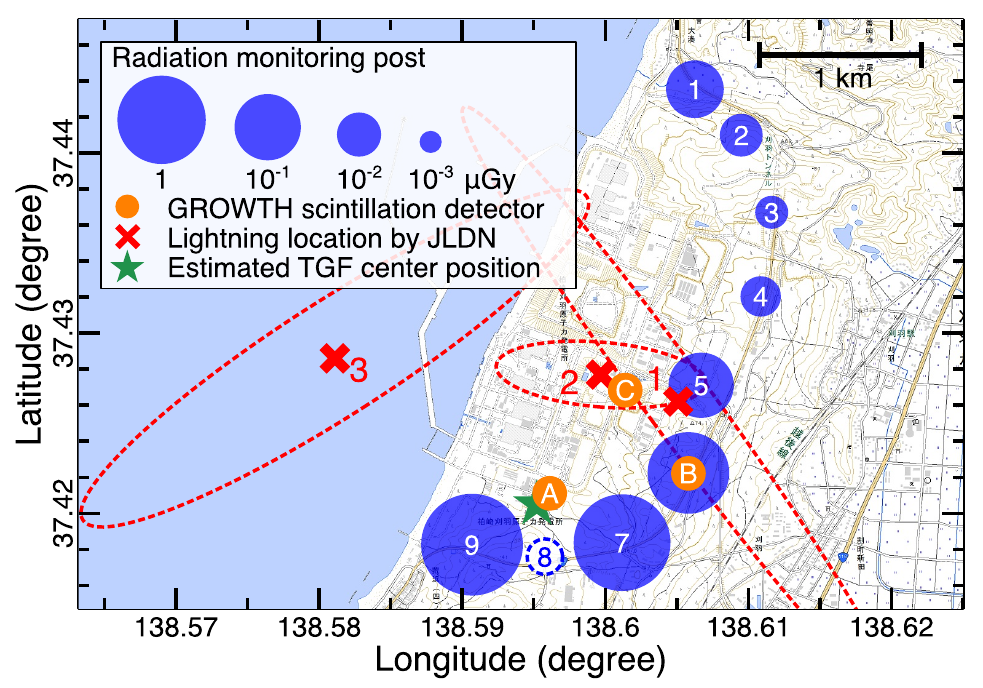}
	\caption{
	Locations of the detectors in Kashiwazaki-Kariwa Nuclear Power Station. 
	Orange circles show Detectors A to C, blue circles monitoring posts (MP), 
	and green star the estimated position where the TGF took place. 
	The size of the blue circles indicates dose of the present TGF. 
	Red crosses and dashed-line ellipses show 
	positions of lightning discharges and their errors reported by JLDN, respectively.
	The numbers 1--3 indicate the time order of these discharges, as shown in Fig.2(h).
	Detector~B and MP6 are installed cospatial.
	The MP8 data were unavailable in the present study.
	}
	\end{center}
\end{figure}

Detectors A, B, and C employ a ${\rm Bi}_{4}{\rm Ge}_{3}{\rm O}_{12}$ (BGO) crystal scintillator of 25.0$\times$8.0$\times$2.5~cm$^{3}$ each,
	and are sensitive to 0.2--18.0~MeV, 0.2--26.0~MeV, and 0.3--15.0~MeV gamma rays, respectively.
	Scintillation pulses from these BGO crystals are read by phototubes. 
	Analog outputs from each phototube are processed by a charge amplifier and a shaping amplifier, 
	with time constants of $10~\mu{\rm s}$ and $2~\mu{\rm s}$, respectively.
	The amplified analog waveforms are sampled by our original data acquisition (DAQ) system 
	with 50~MHz analog-to-digital convertors\cite{Enoto_2017,Wada_2018}.
	Once the DAQ system detects a pulse, it stores the digitized pulse waveform during a $20~\mu{\rm s}$ gate time,
	and measures its maximum and minimum values.
	The maximum value, corresponding to the pulse height, is converted 
	into the deposited energy in the crystal via energy calibration.
	Minimum values present baseline of the analog outputs, and normally remain $\sim$0~volt. 
	The DAQ system can count pulses up to 10~kHz properly.
	The absolute timing of the detectors is adjusted by global-positioning-system signals 
	to an accuracy of $<1~\mu{\rm s}$.
	
In addition, radiation monitoring posts (MPs), operated by the power station and distributed as in Fig.1, 
	measure dose rates every 30 seconds.
	Each station consists of a $\phi5.1~{\rm cm}~\times~5.1~{\rm cm}$ NaI scintillation detector (NaI), and an ionization chamber (IC),
	a sphere of 2-mm thick stainless steal filled with 14-L argon gas at 4 atms. 
	These NaIs and ICs are sensitive to gamma rays of 0.05--3.0~MeV and $>0.05~{\rm MeV}$, respectively.
	While NaIs are dedicated to low-dose-rate measurements up to 10~$\mu{\rm Gy}~{\rm h}^{-1}$,
	ICs stand high rates up to $100~{\rm mGy}~{\rm h}^{-1}$.
		
Lightning activities are monitored by a broadband low-frequency (LF) radio receiver which we installed at Nyuzen 
	(36.954$^{\circ}$N, 137.498$^{\circ}$E), 110~km west-southwest from the observation site.
	The LF receiver consists of a flat-plate antenna sensitive to the 0.8--500~kHz band. 
	Analog outputs from the antenna are sampled by a 4~MHz digitizer\cite{Takayanagi_2013}.
	We also utilize commercial information of Japan Lightning Detection Network (JLDN),
	to obtain position, timing, discharge current, and classification (cloud-to-ground or in-cloud) of lightning pulses.

\section{Results}
During a thunderstorm on 2017 November 24, we observed a lightning discharge 
	and a radiation burst simultaneously, at 10:03:02.282827~UTC, 
	which we hereafter employ as the origin of the elapsed time.
	Figure~2(a)--(c) present time series of count rates obtained by Detectors A--C, respectively, 
	and Fig.~2(d) an LF waveform which continued for $\sim 400~{\rm ms}$.
	The gamma-ray burst had a steep onset ($<10~{\rm ms}$) coincident with the lightning discharge, 
	followed by an exponential decay with a time constant of 59, 47 and 48~ms for Detectors A, B, and C, respectively.
	As shown in Enoto et al.\cite{Enoto_2017}, these decay constants are consistent 
	with those of neutron thermalization, and subsequent emission of de-excitation gamma rays
	through neutron captures (mainly $^{14}{\rm N}~+~{\rm n}~\to~^{15}{\rm N}~+~\gamma$). 
	In addition, Detector~A detected an afterglow, 
	lasting for $\sim 10~{\rm sec}$ and mainly consisting of 511-keV annihilation photons.
	This is due to positrons emitted from $\beta^{+}$-decaying nuclei such as $^{13}{\rm N}$\cite{Enoto_2017}. 
	Therefore, we infer that photonuclear reactions such as $^{14}{\rm N}~+~\gamma~\to~^{13}{\rm N}~+~{\rm n}$ 
	occured following the lightning discharge.
	
\begin{figure}[t]
	\begin{center}
	\includegraphics[width=\hsize]{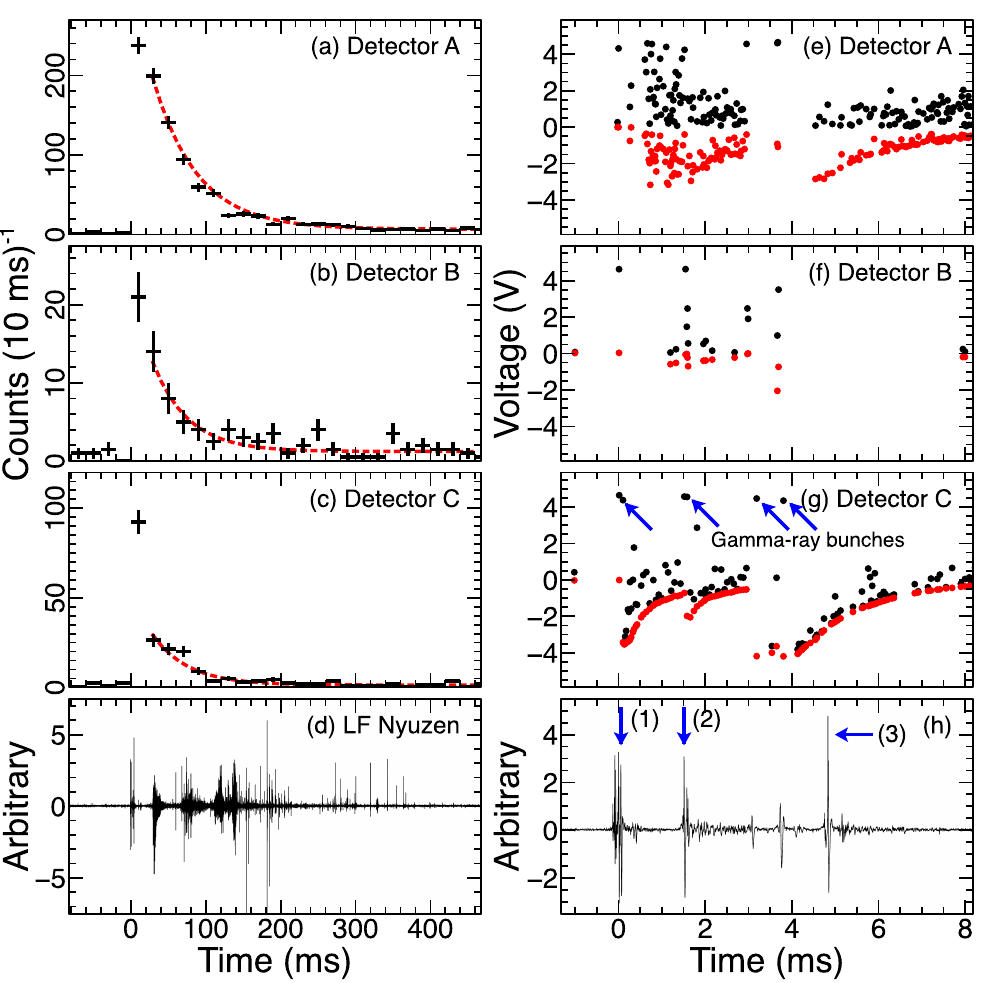}
	\end{center}
	\caption{
	(a)--(c) Count-rate histories of Detectors A--C with 10-ms bins. 
	Overlaid red lines show the best-fit exponential functions.
	(d) An LF waveform obtained by the Nyuzen station, corrected for the propagation time from the station to the detector site.
	(e)--(g) An expanded time series over $-1<t<8~{\rm ms}$. 
	Each gamma-ray photon is represented
	by a pair of black and red dots, which means the maximum and minimum analog voltage 
	from the shaping amplifier output (see text). Blue arrows in (g) show the four gamma-ray bunches.
	(h) An expansion of panel (d). Timings of pulses reported by JLDN are presented by blue arrows.
	The number of pulses corresponds to that in Fig.1.
	}
\end{figure}

Figure~2(e)--(g) expand the initial phase of the gamma-ray burst, where the maximum and minimum values 
    	of the analog outputs measured every 20~$\mu$s are plotted in photon-by-photon basis. 
	During $0<t<5~{\rm ms}$, the maximum values of several photons 
	exceeded the saturation level of the analog amplifier ($\sim$4~volts).
	At the same time, the minimum values went significantly negative,
	and gradually returned to their normal baseline level.
	This behavior, called ``baseline undershoot'', is most clearly seen in Fig.2(e) and (g).
    	These saturated pulses and baseline undershoots are caused 
    	when large charge pulses from the phototube are fed into the charge amplifier.
	As in\cite{Enoto_2017}, this provides clear evidence of a downward TGF, 
	namely, large energy deposits into the BGO crystals in $<$1~ms.
	
In Fig.2(g), the saturated signals of Detector~C, accompanied by the undershoots,
    	can be recognized on four occasions, at $t \sim 0.0$, 1.5, 3.1, and 3.7~ms.
    	The same effects, though less obvious, are seen by Detectors A and B.
    	Therefore, the downward TGF is considered to consist of four gamma-ray ``bunches'' with a duration of $<$1~ms each,
    	and each of them reached our three detectors simultaneously.
	The interval between adjacent bunches ranges over 0.71--1.52~ms, on average 1.22~ms.
	
Although these bunches must contain a large number of gamma-ray photons,
	our DAQ system can record only one photon event during each 20~$\mu{\rm s}$ gate time, 
	even if multiple photons arrive in this period.
    	Furthermore, once saturated, our charge amplifiers take $\sim$1~ms 
    	to recover from the negative undershoot (Fig.2g).
    	The forepart of this interval becomes a dead time, in which subsequent events cannot be acquired.
    	Thus, our detectors were not able to resolve these bunches into individual photons.
    	Instead, we can use the saturated signals as the onset of a gamma-ray bunch.
    	As the negative undershoot becomes $\geq-3$~volts, the detectors resume detecting the photons
    	(mainly from the neutron capture processes), but their pulse heights can be incorrect 
    	by at least to $t \sim 10$~ms.

Figure~2(h) expands the LF waveform.
    	Over $0 < t < 6~{\rm ms}$, when the bunches were detected, we notice 5 to 6 LF spikes.
    	The highest three of them are coincident with discharges reported by JLDN, 
    	a cloud-to-ground pulse of $+10~{\rm kA}$ at $t \sim 0.06~{\rm ms}$, 
	an in-cloud pulse ($-7$~kA, $t \sim 1.51~{\rm ms}$), 
	and a cloud-to-ground pulse ($-10$~kA, $t \sim 4.82~{\rm ms}$),
	indicated as (1), (2), and (3) in Fig.2(h), respectively.
	Their positions reported by JLDN are shown in Fig.1.
	The first two of them coincide with the first two gamma-ray bunches,
	and furthermore, two smaller LF pulses at $t \sim 3.1$ and 3.7~ms, though too faint for JLDN,
	are approximately coincident with the third and fourth bunches.
    	In contrast, we observed no gamma-ray bunches coincident with the third JLDN-reported pulse.

To examine the suggested association between the LF pulses and the gamma-ray bunches,
    	we further expand, in Fig~3, the five pulses of Fig.2(h), 
	where the onset timing of the four gamma-ray bunches are also displayed.
    	The first bunch at $t = 0.0~{\rm ms}$ took place with a positive LF pulse, 
    	which followed a train of negative-polarity fast pulses in $-0.14 < t < 0.0~{\rm ms}$, namely stepped leaders,
    	and preceded the return stroke at $t=0.58$~ms which JLDN reported.
	The second bunch coincides with an JLDN-reported in-cloud pulse.
	The third and forth bunches are also correlated with LF pulses which have bipolar shapes.
	However the onset timings of these bunches have larger uncertainties by 0.1--0.2~ms,
	which are caused probably because the trigger timing of the subsequent bunch 
    	is affected by the charge amplifier recovery from the preceding bunch.
	
\begin{figure}[t]
	\begin{center}
	\includegraphics[width=\hsize]{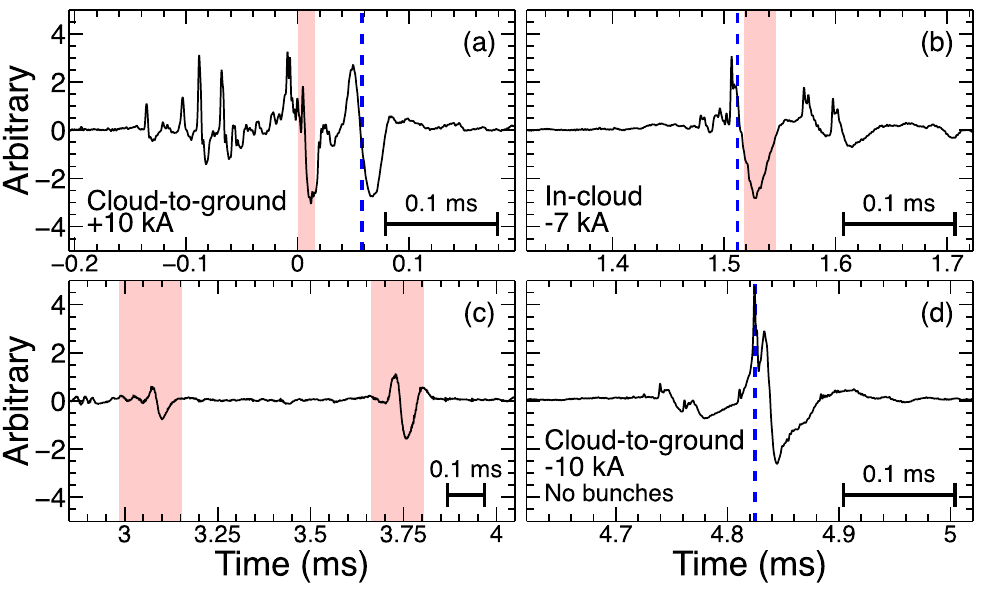}
	\caption{
	Expanded LF waveforms of Fig.2(h). 
	Timing of the lightning discharges reported by JLDN are indicated with blue-dashed lines. 
	The estimated onset timing of each gamma-ray bunch is shown by the red-shaded region, 
	of which the width reflects differences among the three scintillators.
	As customary with atmospheric electricity, pulses with a positive onset indicate negative polarity, 
	namely downward negative current or upward positive current.
	}
	\end{center}
\end{figure}

Besides Detectors A--C, eight ICs also detected the entire burst, in the form of a clear increase in their 30-sec dose rates.
    	In particular, those at MP9 and MP6 reached 170 and 25~$\mu{\rm Gy}$~h$^{-1}$, respectively.
	After subtracting the environmental background, the IC doses integrated for 30~sec become 1.4 and 0.2~$\mu{\rm Gy}$, respectively (Fig.1).
	Assuming a bunch duration as $<$1~ms ($<$4~ms in total), the instantaneous dose rates reach 1.3 and 0.2 Gy~h$^{-1}$, respectively.
    	The data of MP8 were unavailable because it temporarily went down just after the lightning discharge.
    	Since ICs have tolerance for high gamma-ray fluxes by measuring accumulated doses instead of counting photons,
	their increases must be contributed by both the downward TGF (the 4 bunches) and the sub-second emission component.

A slight dose increase, by $\sim10^{-4}~\mu{\rm Gy}$ or less, was also detected by NaIs.
    	Being typical scintillation detectors, they also failed to properly count gamma rays in the bunches,
	and hence have measured doses only of the sub-second component, 
	which originates from neutron captures\cite{Enoto_2017} and contains multiple emission lines up to 10.8~MeV. 
	By extrapolating from the 0.05--3.0~MeV NaI doses, those up to 10.8~MeV were estimated as 
	$2 \times 10^{-3}~\mu{\rm Gy}$ at MP9, which is $<$1\% of the actually measured IC doses.
	Therefore, we conclude that the IC dose increases were caused predominantly by the downward TGF, 
	rather than by the sub-second emission from neutrons. In the same way, doses by the direct neutrons are negligible.
	The present gamma-ray burst, as a whole, is inferred to have involved 
	$>$100~times larger number of gamma-ray photons than those apparently recorded in Fig.2(e)--(g).

\section{Monte Carlo Simulations}
To quantify the present downward TGF, 
	we performed Monte Carlo simulations with {\tt Geant4}\cite{Agostinelli_2003}.
	As a preliminary attempt, we irradiated a mass model of IC with 0.05--50~MeV gamma rays,
	which follow a spectrum as $E^{-1} \exp(-E/7.3~{\rm MeV})$, with $E$ being the gamma-ray energy,
	as expected for bremsstrahlung emission by avalanche electrons\cite{Dwyer_2012b}.
	Then, a typical IC dose, 0.050~$\mu{\rm Gy}$, was found to correspond to a gamma-ray fluence of $10^{4}$~photons~cm$^{-2}$.
	This calibration gives the fluences at MP9 and MP6 as $2.8 \times 10^{5}$ and $4.2 \times 10^{4}~{\rm photons}~{\rm cm}^{-2}$, respectively.
	The present TGF spectrum, although not measured by our detectors, 
	could have had an additional power-law component at higher energies\cite{Tavani_2011,Celestin_2012}.
	Its inclusion into this simulation, with parameters fixed to those of Tavani et al.\citep{Tavani_2011}, 
	was found to increase the fluences necessary to explain the IC doses by only $\sim$30\% or less.
	Further assuming the total TGF duration conservatively as $<10$~ms, 
	the gamma-ray fluxes which arrived at MP9 and MP6 are estimated as 
	$> 2.8 \times 10^{7}$  and $>0.4 \times 10^{7}~{\rm photons}~{\rm cm}^{-2}~{\rm s}^{-1}$, respectively.
	Although similar fluxes must have arrived at Detectors A--C, the peak gamma-ray fluxes they actually measured
	are at most $\sim 300~{\rm photons}~{\rm cm}^{-2}~{\rm s}^{-1}$ (from their effective areas and Fig.1).
	Thus, we conclude that the present TGF had a gamma-ray flux which is $10^4$ to $10^5$ times higher than 
	those apparently recorded by Detectors A--C, and completely saturated them.

Then, we performed full simulations starting from electrons.
    	A two-stage simulation was employed for computational economy. 
	In the first stage, initial particles, namely, avalanche electrons were injected in a mass model of the atmosphere,
	to simulate their atmospheric interactions and track their bremsstrahlung gamma rays. 
	Information of the gamma rays, such as energy, position and momentum vector was collected when they reached the ground.
	In the second stage, the gamma rays reaching the ground were input into the IC mass model, like in the preliminary examination.

In the above simulation, the initial electrons were assumed to follow the theoretical spectrum of RREA, 
	$\exp(-\epsilon/7.3~{\rm MeV})$\cite{Dwyer_2011}, where $\epsilon$ is the electron energy.
	The electron beam is assumed to be narrow and downward, with neither a tilt angle nor beam divergence.
	At an altitude from 1000~m to 3000~m with a 250-m step, $10^{9}$ electrons of 1--50~MeV were injected.
	Initial electrons below 1~MeV were ignored because their products are mostly absorbed before reaching the ground.
	The left panel of Fig.4 shows the results from these simulations, in the form of lateral spread of gamma-ray doses, 
	as caused by struggling of electrons and scatterings of bremsstrahlung photons.

\begin{figure}[bth]
\begin{center}
\includegraphics[width=\hsize]{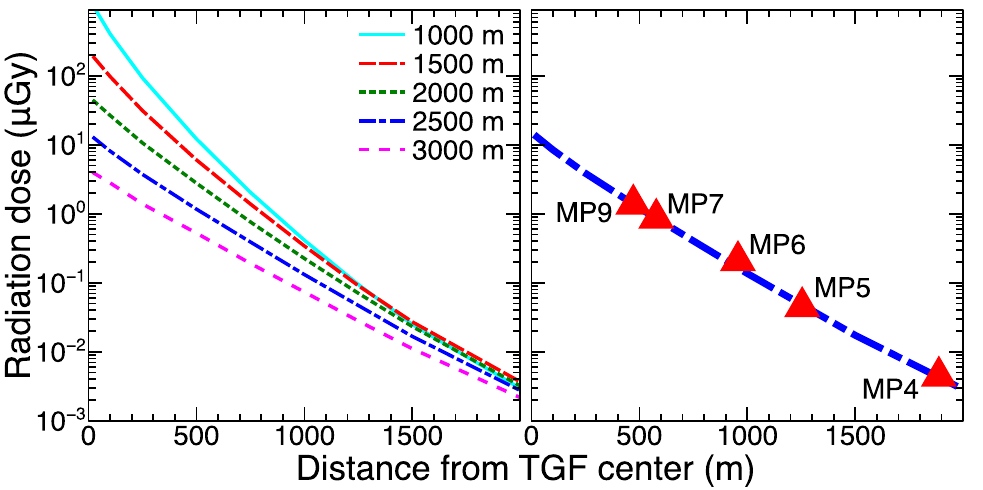}
	\caption{
	Results of Monte Carlo simulations. 
	(left) Air absorbed doses shown as a function of the offset from the TGF center.
	Different colors specify different initial height of injected electrons.
	The obtained doses are normalized to an electron number of $8 \times 10^{18}$ above 1~MeV.
	(right) The best-fitting model to the IC data (see text).
	}
\end{center}
\end{figure}

The Monte-Carlo predicted doses at each altitude were compared with the IC measurements at different locations.
    	By changing the TGF center position with the 50-m step,
    	and using the IC data (including 15\% systematic errors) within 2000 m of the assumed TGF center, 
    	we searched for the best fit solution via chi-square evaluation.
    	Then, as shown in the right panel of Fig.4, the fitting has become acceptable 
    	with $\chi^{2}=2.19$ for 1 degree of freedom 
	(5~data points and 4 parameters, namely 2-dimensional position, altitude and normalization), 
    	for the total avalanche electrons above 1~MeV of $8^{+8}_{-4} \times 10^{18}$, 
    	the TGF height of $2500\pm500~{\rm m}$ altitude, and the TGF center being 100 m southwest from Detector~A (Fig.1). 
	Errors are at a 1~$\sigma$ confidence level.
	The on-ground fluences of $>$1~MeV gamma rays from the four bunches 
	are estimated as $4 \times 10^{5}~{\rm photons}~{\rm cm}^{-2}$ at the TGF center, 
	and $7 \times 10^{3}~{\rm photons}~{\rm cm}^{-2}$ at an offset of 1~km.

\section{Discussion}
The present burst can be understood as a downward TGF, which consisted of the four gamma-ray bunches 
    	like multi-pulse TGFs observed from space\cite{Foley_2014}.
	The intervals between adjacent bunches in the present case, a few millisecond, 
	are similar to those of multi-pulse TGFs\cite{Fishman_1994,Foley_2014}.	
	As seen in Fig.3(a), the first bunch took place at the last phase of the leader development,
	in agreement with several  previous reports\cite{Lu_2010,Cummer_2015,Abbasi_2018,Smith_2018}.
	Since the LF pulses for $-0.15 < t <0$ ms are negative, the relevant leaders must be downward negative ones or upward positive ones.
	In either case, electrons should be accelerated downward, as required for the occurrence of  downward TGFs.
	In addition, the first bunch clearly preceded the return stroke.
	This agrees with a simple physical expectation that electrostatic acceleration should be hampered in return strokes,
	due to azimuthal magnetic filed created by the strong electric current.
	The other bunches, though coincident with negative LF pulses, 
	also have less clear relations to return strokes of cloud-to-ground currents.
	
Ordinary TGFs, observed from space, typically initiate at altitudes of $\geq$8~km\cite{Cummer_2015,Mailyan_2016},
	namely in middle to upper layers of summertime thunderclouds.
	In contrast, the estimated altitude of 2.5$\pm$0.5~km in the present case is much lower, 
	where the atmosphere is approximately twice denser.
	Since winter thunderstorms in Japan typically have $\sim$5~km cloud tops and $<$1~km cloud bases\cite{Kitagawa_1992},
	the present TGF is likely to have taken place in a lower to middle layer of the winter thunderclouds, 

Using the RHESSI TGF data, Dwyer and Smith\cite{Dwyer_2005c} estimated the number of involved electrons ($>$~1~MeV) 
	as $1 \times 10^{16}$--$2 \times 10^{17}$, 
	and Mailyan et al.\cite{Mailyan_2016}, using inidividual Fermi TGF data, 
	derived as $4 \times 10^{16}$--$3 \times 10^{19}$, with an average of $2 \times 10^{18}$.
	The estimated electron number of the present burst is in the range of Mailyan et al.
	Therefore, the downward TGF in the present case have approximately the same properties as ordinary TGFs, 
	despite the atmospheric density difference.

During winter thunderstorms in Japan, Bowers et al.\cite{Bowers_2017} detected a downward TGF, 
	and indirectly estimated its total number of photons above 1~MeV as $\sim10^{17}$, 
	considering that it produced neutrons via photonuclear reactions.
	The electron number above 1~MeV of our estimation, $8^{+8}_{-4} \times10^{18}$, 
	can be converted into the initial number of bremsstrahlung photons as $7^{+7}_{-3}~\times~10^{17}$.
	Therefore, the case of Bowers et al. is similar to ours.
	Furthermore, we succeeded for the first time in directly measuring the doses included in a downward TGF 
	that triggered photonuclear reactions, and in evaluating the number of avalanche electrons.
	At the Kashiwazaki-Kariwa site, we previously observed another gamma-ray burst that involved photonuclear reactions\cite{Enoto_2017}.
	We infer that this burst is similar to the present one and that of Bowers et al.\cite{Bowers_2017},
	because the number of photoneutrons in this case is estimated to be of the same order as that of Bowers et al.

In summary, we observed, during a winter thunderstorm, a downward TGF consisting of four gamma-ray bunches coincident with LF pulses. 
	The IC doses and the Monte Carlo simulations allowed us to estimate the total number of avalanche electrons
	as $8 \times 10^{18}$ above 1~MeV, produced at the 2500~m altitude. 
	The present result suggests that downward TGFs in winter thunderclouds have characteristics very similar to those of up-going TGFs, 
	except the electrons being accelerated in the opposite directions into much thicker atmosphere.

\begin{acknowledgements}
We thank the radiation safety group of Kashiwazaki-Kariwa Nuclear Power Station of Tokyo Electric Power Company Holdings
	for providing the observation site and the MP data.
	The BGO scintillation crystals were provided by Sakurai Radioactive Isotope Physics Laboratory of RIKEN Nishina Center.
	We appreciate discussions with T. Torii, M. Ikuta, A. Bamba, H. Odaka, T. Tamagawa, and advice for simulations by G. S. Bowers.
	The Monte Carlo simulations were performed on the HOKUSAI BigWaterfall supercomputing system 
	of RIKEN Advanced Center for Computing and Communication.
	The deployment of the LF receiver was supported by Town of Nyuzen, Toyama Prefecture.
	This research is supported by JSPS/MEXT KAKENHI grants 15K05115, 15H03653, 16H06006, 16K05555, 18J13355, 18H01236, 19H00683, 
	by SPIRITS 2017 and Hakubi projects of Kyoto University, by the Hayakawa Satio Fund of the Astronomical Society of Japan,  
	and by citizen supporters via an academic crowdfunding platform ``academist''. 
	The background image in Fig.1 was provided by the Geospatial Information Authority of Japan.
\end{acknowledgements}



%



\end{document}